\documentclass[aps,prd,twocolumn,nofootinbib]{revtex4}
\usepackage{graphicx}
\usepackage{graphics}
\usepackage{bbm}
\usepackage{longtable}

\begin{document}


\title{Quantum Dissipation and CP Violation in MINOS}


\author{R.L.N. Oliveira}
\email{robertol@ifi.unicamp.br}
\affiliation{Instituto de F\'\i sica Gleb Wataghin\\
Universidade Estadual de Campinas, UNICAMP\\
13083-970, Campinas, S\~ao Paulo, Brasil}

\author{M.M. Guzzo}
\email{guzzo@ifi.unicamp.br}
\affiliation{Instituto de F\'\i sica Gleb Wataghin\\
Universidade Estadual de Campinas, UNICAMP\\
13083-970, Campinas, S\~ao Paulo, Brasil}

\author{P.C. de Holanda}
\email{holanda@ifi.unicamp.br}
\affiliation{Instituto de F\'\i sica Gleb Wataghin\\
Universidade Estadual de Campinas, UNICAMP\\
13083-970, Campinas, S\~ao Paulo, Brasil}



\date{\today}

\begin{abstract}

We use the open quantum systems framework to analyze the MINOS data and perform this analysis considering two different dissipative models. In the first model, the dissipative parameter describes decoherence effect and in the second, the dissipative parameter describes other dissipative effects including decoherence. With the second model it is possible to study CP violation since we consider Majorana neutrinos. The analysis from the muon neutrino and antineutrino beam assigns different values to all the parameters of the models, but consistent with each other. Assuming that neutrinos are equivalent to antineutrinos, the global analysis presents nonvanishing Majorana CP phase depending on the energetic parameterization of the dissipative parameter.

\end{abstract}


\maketitle

\section{Introduction}

The open quantum system can be used in neutrino physics in order to study the dissipative effects and the oscillation phenomena \cite{workneut,ben}. In general, one can use the Lindblad master equation to describe the neutrino beam evolution, where together with the oscillation parameters, new parameters arise and indicate how the dissipative effects act in this system \cite{lin,gor,dav,len}. 

Currently, there are some important results in neutrino oscillations, for example, the determination of  $\theta_{13}$ mixing angle and the results obtained from neutrino and antineutrino beam by MINOS \cite{day,double,mimi2,mimi3}. As it is well known, MINOS is a long base line experiment where the flux of neutrino peaks at $3$ GeV and its beam is mainly characterized by oscillations between $\nu_{\mu}\leftrightarrow\nu_{\tau}$ (or $\bar{\nu}_{\mu}\leftrightarrow\bar{\nu}_{\tau}$) \cite{1min, mimi1, mimi2}.  In especial, when we treat oscillation in vacuum, the Lindblad master equation has simple form and its application is direct \cite{workneut,ben}. If we only assume oscillation between $\nu_{\mu}\leftrightarrow\nu_{\tau}$ (or $\bar{\nu}_{\mu}\leftrightarrow\bar{\nu}_{\tau}$), the Lindblad master equation is easily adapted to study the MINOS experiment and this framework do not need to be modified because, in this case, the effective matter potential is not important.

Many models are obtained from the Lindblad master equation when it is used to study neutrino oscillation in vacumm \cite{workneut}. Notoriously, the model with decoherence effect is the only one of the seven models that adds only one parameter in the oscillation pattern that has really been studied until now \cite{dea, lis, fo, liss, mimi1, dan, yu, mmy, dan, fun1, gab}. All these seven models satisfy the complete positivity \cite{workneut,dav,lin,len}. However, it is clear that there are other models very interesting \cite{workneut}. Here, we present a data analysis from the MINOS experiment where we use two dissipative models and also the standard oscillation model. 

The analysis with standard oscillation model is introduced in order to verify if our simple approach is enough to understand MINOS results. Then, we present the analysis using the first dissipative model that adds decoherence in the neutrino oscillation and after, we introduce the analysis using  the second dissipative model that includes other dissipative effects in addition to decoherence effect. Interest enough, if we consider Majorana neutrinos, the second model presents a dependence on the CP phase in its survival probability even in two families.

Our results show that the analysis from the muon neutrino and antineutrino beam assigns different values to all the parameters of the all models \cite{mimi2,mimi3,chun}. However, presently there is consistence between these values and CPT violation seems unlikely. Then, assuming neutrinos are equivalent to antineutrinos, we present the global analysis and, depending on the energetic parameterization of the dissipative parameter, the Majorana CP phase has a value non-zero.

In the course of the present study, we show that the second model fits very well the MINOS data and in some cases, even assuming that $\nu_{\mu}$  is equivalent to $\bar{\nu}_{\mu}$, CP violation can occur depending on energetic parameterization of the dissipative parameter and the oscillation  probabilities of neutrino and antineutrino are always different from each other.  

\section{Formalism}

Quantum dissipation occurs in all quantum system and when any quantum system is written as a state superposition, the dissipation effects become more evident. A well-known example of this effect is decoherence, but there are other important dissipative effects. From the MINOS data we want to quantify and bound some of these quantum dissipative effects. We follow the approach introduced by reference \cite{workneut}, where only one more parameter was included in neutrino oscillation theory. In particular, we are interested in two specific models. The two family neutrino survival probabilities of these models are written as
\begin{equation}
P^{C1}_{\nu_{\mu}\rightarrow\nu_{\mu}}	 = 1 - \frac{1}{2}\sin^{2}(2\theta)\Bigg[1-e^{-\gamma_{0} x}\cos\bigg(\frac{\Delta m^{2}}{2E} x\bigg)\Bigg]\,
\label{i.i}
\end{equation}
and
\begin{eqnarray}
\tilde{P}^{C7}_{\nu_{\mu}\rightarrow\nu_{\mu}} & = &\frac{1}{2}+e^{-\gamma_{0} x}\Bigg\{\frac{1}{2}-\sin^{2}(2\theta)\sin^{2}\bigg(\frac{\Delta m^{2}}{4E} x\bigg)\nonumber \\ & &+ \frac{\gamma_{0} E}{2\Delta m^{2}}\sin\phi\sin(4\theta)\sin\bigg(\frac{\Delta m^{2}}{2E} x\bigg)\Bigg\}\,,\nonumber \\ & &
\label{i.ii}
\end{eqnarray}
where $\Delta m^{2}=m^{2}_{3}-m^{2}_{3}$ is the mass square difference, $\theta$ is the mixing angle, $\gamma_{0}$ is the dissipative effect and $x$ is the distance between the source and the detector. Note that the $\gamma_{0}$ parameter has different meaning in Eqs. (\ref{i.i}) and (\ref{i.ii}). In Eq. (\ref{i.i}), the $\gamma_{0}$ describes decoherence and in Eq. (\ref{i.ii}) describes a more general quantum dissipative effect, as it was discussed in the reference \cite{workneut}. Furthermore, we are following the same notation of \cite{workneut}, where superscript \textit{Case 1} and \textit{Case 7} refers to \textit{Case 1} and \textit{Case 7} which were analysed in Ref. \cite{workneut} and $\tilde{P}$ means that survival probability is obtained when $\gamma_{0}^{n}\rightarrow 0$ to $n\geq2$. 

As it is usual, we will assume an energy dependence of $\gamma_{0}$ by means of a power-law written as
\begin{eqnarray}
\gamma_{0} & = & \gamma\left(\frac{E}{E_{0}}\right)^{n}\,,
\label{i.iii}
\end{eqnarray}
where $n=0,\pm 1, \pm 2$. The energy scale, $E_{0}$, modulates the magnitude expected for the dissipation effects. This procedure is performed because the effects are included phenomenologically and, in the present moment, it is not possible to determine if these effects are due to quantum gravity \cite{ell,haw} or to a hypothetic medium with reservoir behavior as it is thought via open quantum system approach \cite{pet,dav,uri,len,joo}. 

We also consider de usual survival probability that can be obtained directly from Eq. (\ref{i.i}) and (\ref{i.ii}) when we lead $\gamma_{0}\rightarrow 0$, thus, it is written as 
\begin{eqnarray}
P_{\nu_{\mu}\rightarrow\nu_{\mu}} & = &1-\sin^{2}(2\theta)\sin^{2}\bigg(\frac{\Delta m^{2}}{4E} x\bigg)\,.
\label{i.iv}
\end{eqnarray}

In order to clearly see the dissipative effects acting in the neutrino propagation, we will use a very simple approach to perform the analysis. We will use only the ratio to no oscillation that can be obtained supposing \cite{kim}
\begin{equation}
P_{\nu_{\mu}\rightarrow\nu_{\mu}}=\frac{N^{obs}_{\nu_{\mu}}}{N^{no\mbox{\small{-}}osc}_{\nu_{\mu}}}\,,
\label{i.v}
\end{equation}
where $N^{obs}_{\nu_{\mu}}$ and $N^{no\mbox{\small{-}}osc}_{\nu_{\mu}}$ are, respectively, the number of observed $\nu_{\mu}$ events and the number of expected $\nu_{\mu}$ events in the absence of oscillations. From the muon neutrino beam, we assume the ratio to no oscillation that can be obtained by mean of the article \cite{mimi1} in which we take the superior error bar as the probability uncertainty. The ratio to no oscillation from the muon antineutrino beam are obtained in the article \cite{mimi3}, where using the Eq. (\ref{i.v}) we find this ratio and define the probability uncertainty as
\begin{equation}
\Delta P_{\nu_{\mu}\rightarrow\nu_{\mu}}=\frac{\sqrt{N^{obs}_{\nu_{\mu}}}+\alpha}{N^{no\mbox{\small{-}}osc}_{\nu_{\mu}}}\,.
\label{i.vi}
\end{equation}

In this case, we assume also that the superior error bars are $ N^{obs}$ uncertainty. So, $\alpha$ is a factor that reflects the systematic uncertainty obtained by mean of the difference between $N^{obs}_{\nu_{\mu}}$ data uncertainty and $N^{obs}_{\nu_{\mu}}$ statistic uncertainty. 

In order to improve our analysis, we calculate the mean value of the survival probabilities, Eq. (\ref{i.i}),(\ref{i.ii}) and (\ref{i.iv}), in each range energy where was it defined a bin energy. Furthermore, we consider, for sake of simplicity, the following definition for $\chi^{2}$ function
\begin{equation}
\chi^{2}=\sum_{i} \frac{\left(P^{i}_{exp}-P^{i}_{theo}\right)^{2}}{\sigma^{2}_{i}}
\label{i.vii}
\end{equation}
where $P^{i}_{exp}$ is the data obtained using the Eq. (\ref{i.vi}), $P^{i}_{theo}$ is the theoretical survival probability  and $\sigma$ is the uncertainty defined in Eq. (\ref{i.vi}). We also define the global $\chi^{2}_{glob}$ as 
\begin{equation}
\chi^{2}_{glob}= \chi^{2}_{\nu}+\chi^{2}_{\bar{\nu}}\,,
\label{i.viii}
\end{equation}
once that we will take into account in our analysis that neutrinos can be equivalent to antineutrino and the dissipative effect must happen in both channel.

\section{Results and Discussions}

We start the analysis considering the standard oscillation model to verify if the approach introduced before yields results compatible to the MINOS result \cite{mimi1,mimi4}. 
\begin{center}
\begin{table}[!htb]
\begin{ruledtabular}
\caption{\label{tab:table1}The values obtained from the analysis of $\nu_{\mu}$, $\bar{\nu}_{\mu}$ and global hypothesis, $\nu^{g}_{\mu}$. The values to $\nu_{\mu}$ and $\bar{\nu}_{\mu}$ agree at $68\%$ C. L. with MINOS Collaboration \cite{mimi1,mimi4}. }
\begin{tabular}{||c||ccc||}
\hline
Standard&$\nu_{\mu}$&$\bar{\nu}_{\mu}$&$\nu^{g}_{\mu}$\\
\hline
\hline
$\Delta m^{2}(10^{-3}eV^{2})$&$2.34^{+0.09}_{-0.09}$&$2.71^{+0.41}_{-0.53} $&$2.36^{+0.14}_{-0.15} $\\
\hline
$\sin^{2}(2\theta)$&$0.92^{+0.05}_{-0.04}$&$0.94_{-0.16} $&$0.92^{+0.06}_{-0.07} $\\
\hline
$\chi^{2}$&$19.48 $&$19.12$&$39.25 $\\
\hline
\end{tabular}
\end{ruledtabular}
\end{table}
\end{center}
As we can see in the Table \ref{tab:table1}, our results agree at $68\%$ C.L. with the values obtained from MINOS collaboration to both neutrino and antineutrino parameters. MINOS collaboration indicates that the oscillation parameter values are: $\Delta m^{2}=2.32^{+0.12}_{-0.08}$, $\sin^{2}(2\theta)=1.00_{-0.06}$,$\Delta\bar{ m}^{2}=2.62^{+0.40}_{-0.37}$, $\sin^{2}(2\bar{\theta})=0.95^{+0.11}_{-0.12} $\cite{mimi1,mimi4}.

\begin{table*}[!htb]
\begin{center}
\caption{\label{tab:table2}The values obtained from the analysis of $\nu_{\mu}$, $\bar{\nu}_{\mu}$ and global hypothesis, $\nu^{g}_{\mu}$ to the \textit{Case 1} and \textit{Case 7}  models. The oscillation parameter values obtained with \textit{Case 1} model agree at $68\%$ C. L. with the values presented by MINOS Collaboration \cite{mimi1,mimi4}. The values obtained with \textit{Case 7} model have the same agreement with the MINOS Collaboration only when $n>-2$. The superscript asterisk on values of the Majorana CP phase indicates that there is not significant sensitivity for this parameter. }
\begin{ruledtabular}
\begin{tabular}{||c||ccccc||}
\hline
Case 1:$\nu_{\mu}           $&$n=-2                $&$n=-1                 $&$n=0                  $&$n=1                  $&$n=2$\\
\hline
\hline
$\Delta m^{2}(10^{-3}eV^{2})$&$2.30^{+0.19}_{-0.16} $&$2.22^{+0.22}_{-0.09} $&$2.24^{+0.19}_{-0.14} $&$2.27^{+0.17}_{-0.15} $&$2.34^{+0.15}_{-0.16} $\\
\hline
$\sin^{2}(2\theta)          $&$0.95_{-0.09}         $&$1.00_{-0.12}         $&$0.98_{-0.09}         $&$0.96_{-0.07}         $&$0.92_{-0.06} $\\
\hline
$\gamma (10^{-14} eV )      $&$3.72^{+17.81}        $&$7.18^{+7.16}         $&$2.75^{+2.63}_{-2.65} $&$1.20^{+0.45}_{-0.44} $&$0.05^{+0.02}_{-0.02} $\\
\hline
$\chi^{2}                   $&$19.44                $&$18.90                $&$17.64                $&$15.66                $&$17.50 $\\
\hline
\hline
\hline
Case 1:$\bar{\nu}_{\mu}     $&$n=-2                    $&$n=-1                 $&$n=0                  $&$n=1                  $&$n=2$\\
\hline
\hline
$\Delta m^{2}(10^{-3}eV^{2})$&$2.71^{+0.41}_{-0.56}    $&$2.71^{+0.41}_{-0.55} $&$2.70^{+0.37}_{-0.66} $&$2.71^{+0.41}_{-0.53}  $&$2.71^{+0.41}_{-0.53} $\\
\hline
$\sin^{2}(2\theta)          $&$0.94^{}_{-0.16}         $&$0.94_{-0.16}         $&$0.93_{-0.12}         $&$0.94_{-0.16}         $&$0.94_{-0.16} $\\
\hline
$\gamma (10^{-14} eV )      $&$0^{+27.61}              $&$0^{+20.53}           $&$4.02^{+6.70}         $&$0.01^{+0.03}         $&$0.01^{+0.03} $\\
\hline
$\chi^{2}                   $&$19.12                   $&$19.12                $&$18.81                $&$19.06                $&$19.06 $\\
\hline
\hline
\hline
Case 1:$\nu^{g}_{\mu}       $&$n=-2                 $&$n=-1                 $&$n=0                  $&$n=1                  $&$n=2$\\
\hline 
\hline
$\Delta m^{2}(10^{-3}eV^{2})$&$2.32^{+0.19}_{-0.15} $&$2.24^{+0.23}_{-0.09} $&$2.25^{+0.18}_{-0.13} $&$2.36^{+0.15}_{-0.15} $&$2.35^{+0.15}_{-0.15} $\\
\hline 
$\sin^{2}(2\theta)          $&$0.95_{-0.09}         $&$1.00_{-0.13}         $&$0.98_{-0.08}         $&$0.92^{+0.07}_{-0.06} $&$0.92^{+0.06}_{-0.07} $\\
\hline
$\gamma (10^{-14} eV )      $&$3.64^{+16.65}        $&$6.87^{+6.61}         $&$3.10^{+2.37}_{-2.49} $&$0.01^{+0.03}         $&$0.03^{+0.02	} $\\
\hline
$\chi^{2}                   $&$39.21                $&$38.81                $&$37.07                $&$39.18                $&$38.61 $\\
\hline
\end{tabular}
\end{ruledtabular}
\end{center}
\begin{center}
\begin{ruledtabular}
\begin{tabular}{||c||ccccc||}
\hline
Case 7:$\nu_{\mu}            $&$n=-2                $&$n=-1                  $&$n=0                  $&$n=1                  $&$n=2$\\
\hline
\hline
$\Delta m^{2} (10^{-3}eV^{2})$&$8.59^{+0.71}_{-0.61}$&$2.25^{+0.09}_{-0.09}  $&$2.23^{+0.10}_{-0.09} $&$2.28^{+0.09}_{-0.10} $&$2.35^{+0.09}_{-0.10} $\\
\hline
$\sin^{2}(2\theta)           $&$0.95_{-0.10}        $&$0.98_{-0.05}          $&$0.98_{-0.05}         $&$0.95^{+0.05}_{-0.05} $&$0.92^{+0.04}_{-0.05} $\\
\hline
$\gamma (10^{-14} eV )       $&$3.45^{+18.15}       $&$4.67^{+8.71}          $&$2.73^{+2.38}_{-2.18} $&$1.20^{+0.43}_{-0.40} $&$0.04^{+0.02}_{-0.02} $\\
\hline
$\sin^{2}\phi                $&$ 0^{*}              $&$ 0^{*}                $&$0^{*}                $&$ 0^{*}               $&$0^{*}  $\\
\hline
$\chi^{2}                    $&$19.46               $&$19.01                 $&$17.64                $&$15.50                $&$17.45 $\\
\hline
\hline
\hline
Case 7:$\bar{\nu}_{\mu}      $&$n=-2                          $&$n=-1                         $&$n=0                  $&$n=1                  $&$n=2$\\
\hline
\hline
$\Delta m^{2}(10^{-3}eV^{2}) $&$10.09^{+1.53}_{-1.80}        $&$2.71^{+0.41}_{-0.55}          $&$2.70^{+0.37}_{-0.66} $&$2.70^{+0.42}_{-0.52} $&$2.71^{+0.40}_{-0.52} $\\
\hline
$\sin^{2}(2\theta)           $&$0.94_{-0.16}                 $&$0.94_{-0.16}                  $&$0.92_{-0.14}         $&$0.94_{-0.16}         $&$0.94^{+0.08}_{-0.16} $\\
\hline
$\gamma (10^{-14} eV )       $&$1.71^{+27.59}\times10^{-3} $&$1.00^{+20.29}\times10^{-3}  $&$3.71^{+6.96}         $&$0.02^{+0.07}_{-0.01} $&$2.81^{+13.46}_{-1.71}\times10^{-4} $\\
\hline 
$\sin^{2}\phi                $&$0.80^{*}                     $&$ 0^{*}                        $&$ 0^{*}               $&$0.10^{*}             $&$0^{*} $\\
\hline
$\chi^{2}                    $&$19.12                        $&$19.12                         $&$ 18.06               $&$16.87                $&$17.29 $\\
\hline
\hline
\hline
Case 7:$\nu^{g}_{\mu}        $&$n=-2                  $&$n=-1                    $&$n=0                  $&$n=1                  $&$n=2$\\
\hline\hline
$\Delta m^{2} (10^{-3}eV^{2})$&$8.67^{+0.67}_{-0.63}  $&$2.24^{+0.23}_{-0.09}    $&$2.28^{+0.09}_{-0.1}  $&$2.36^{+0.14}_{-0.15}  $&$2.36^{+0.15}_{-0.15} $\\
\hline 
$\sin^{2}(2\theta)           $&$0.94_{-0.08}          $&$1.00_{-0.13}            $&$0.98_{-0.09}         $&$0.92^{+0.07}_{-0.06}  $&$0.92^{+0.07}_{-0.06} $\\
\hline
$\gamma (10^{-14} eV )       $&$3.20^{+17.19}         $&$6.87^{+6.63}            $&$3.66^{+2.42}_{-2.80} $&$0.03^{+0.04}         $&$2.31^{+1.69}_{-1.2}\times10^{-4} $\\
\hline  
$\sin^{2}\phi                $&$0^{*}                 $&$ 1.00^{*}               $&$1.00^{*}             $&$ 0.01^{*}            $&$0^{*} $\\
\hline
$\chi^{2}                    $&$39.23                 $&$38.81                   $&$ 36.85               $&$36.92                $&$37.42 $\\
\hline
\end{tabular}
\end{ruledtabular}
\end{center}
\end{table*}

The analysis from the neutrinos and antineutrinos show consistency \cite{mimi4}, but the values for each parameters are different from each other. Then, we perform a global analysis supposing neutrinos are equivalents to antineutrinos and the results can be seen in the Table \ref{tab:table1}.

As it was expected, the oscillation parameters tend to $\nu_{\mu}$ values when  the global hypothesis is used.

\begin{figure*}[!htb]
\center
\includegraphics[width= 8.1 cm, height=7.5cm]{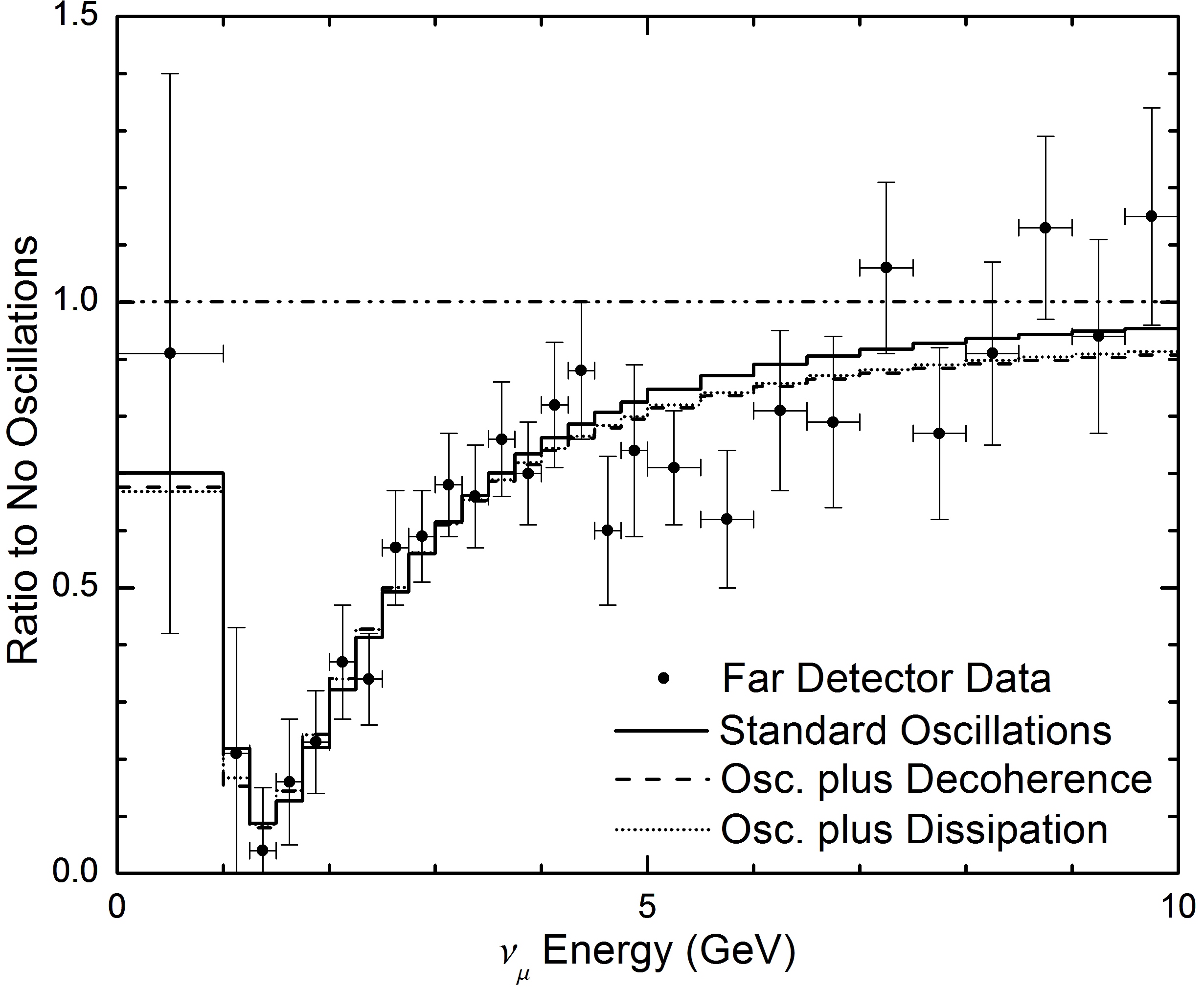}
\includegraphics[width= 8.1 cm, height=7.5cm]{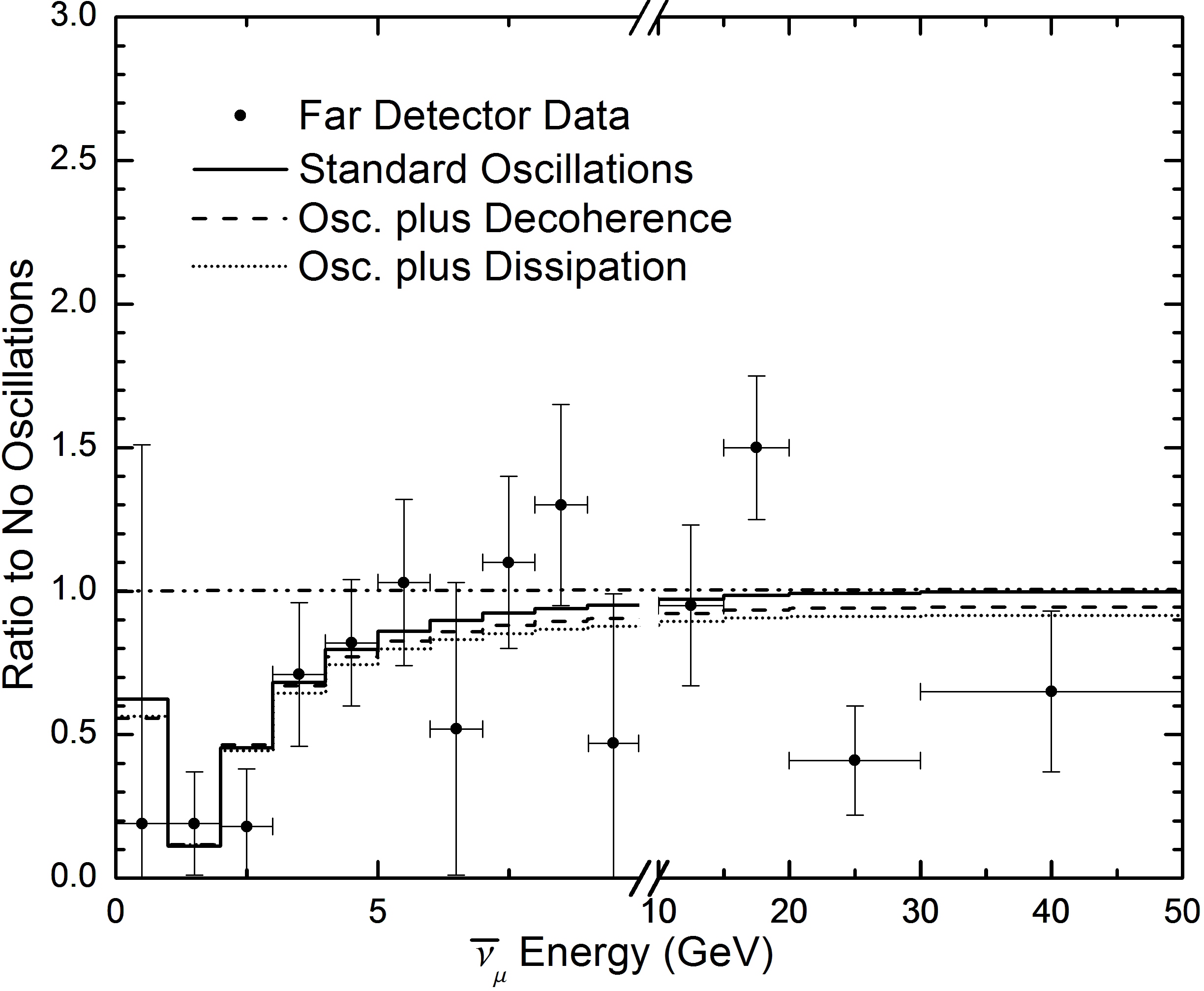}
\caption{The graphics were made using the oscillation parameter values obtained with the equivalence condition between $\nu_{\mu}$ and $\bar{\nu}_{\mu}$. On the left it is shown the neutrino behavior and on the right antineutrino behavior taking $n=0$ in both \textit{Case 1} and \textit{Case 7} models.}
\label{fig.i}
\end{figure*}

Let us include now the dissipative effect in the analysis. We start with the model given by survival probability in Eq. (\ref{i.i}), where decoherence is the dissipative effect coupled in the neutrino oscillation. The results are showed in Table \ref{tab:table2}. The results obtained using the second dissipative model, that includes decoherence and other dissipative effects, can also be seen in Table \ref{tab:table2}.

For all the energy parameterization of $\gamma_{0}$ in the \textit{Case 1} model, the oscillation parameter values remain consistent with each other. This happens also when we compare the oscillation parameter values from the standard oscillation model and \textit{Case 1} model. The same consistency is present between the oscillation parameters in \textit{Case 7} model, but when $n=-2$ in the power-law, the oscillation parameters of the \textit{Case 7} model and standard oscillation model are very different. The $\Delta m^{2}$ in model \textit{Case 7} is greater than in the standard oscillation model, however, it does not change the capacity of this case to fit the data because, in this approach, the important quantity is the mean value of the probability in each bin.

Since we accept our results obtained with the standard oscillation model as being enough to understand the MINOS results, we can conclude that, with exception of the \textit{Case 7} model with $n=-2$, the value of the oscillation parameters obtained in all dissipative cases are consistent with the values obtained from the standard oscillation model.  

From the analysis  for neutrinos and antineutrinos, we can see that the dissipative parameter presented high variance in many cases and all the set of oscillation parameters, i.e., $\Delta m^{2}$ ($\Delta \bar{m}^{2}$) and $\sin^{2}(2\theta)$ ($\sin^{2}(2\bar{\theta})$) in each case,  are different from each other when $n$ varies, but are consistent between neutrinos and antineutrinos in the same case. Furthermore, the results did not present sensitivity to bind the CP Majorana phase and in the most of cases the best fit is $\phi=0$. This panorama shows that there is only a small possibility to happen CPT violation in all models analyzed. Therefore, the equivalence between $\nu_{\mu}$ and $\bar\nu_{\mu}$ behavior is the reasonable hypothesis.         

When we consider this equivalence hypothesis and perform the global analysis, all the models fit the experimental data very well. This can be seen in Fig. \ref{fig.i}, where we plot all the models taking $n=0$ on dissipative models (\textit{Case 1} and \textit{Case 7}). The three lines illustrate the following: the solid line is the behavior of the standard survival probability, dashed line is the behavior of the \textit{Case 1} model and the dot line shows the behavior of the \textit{Case 7} model. On the left (right), we present results for neutrinos (antineutrinos). The behavior of survival probabilities are clear in this energy range and when we treat neutrinos, the larger part of the plot of the \textit{Case 7} model line is above the one of the \textit{Case 1} model. The inverse occurs in antineutrino case.        

\begin{figure*}[!htb]
\center
\includegraphics[width= 5.4 cm, height=5.62cm]{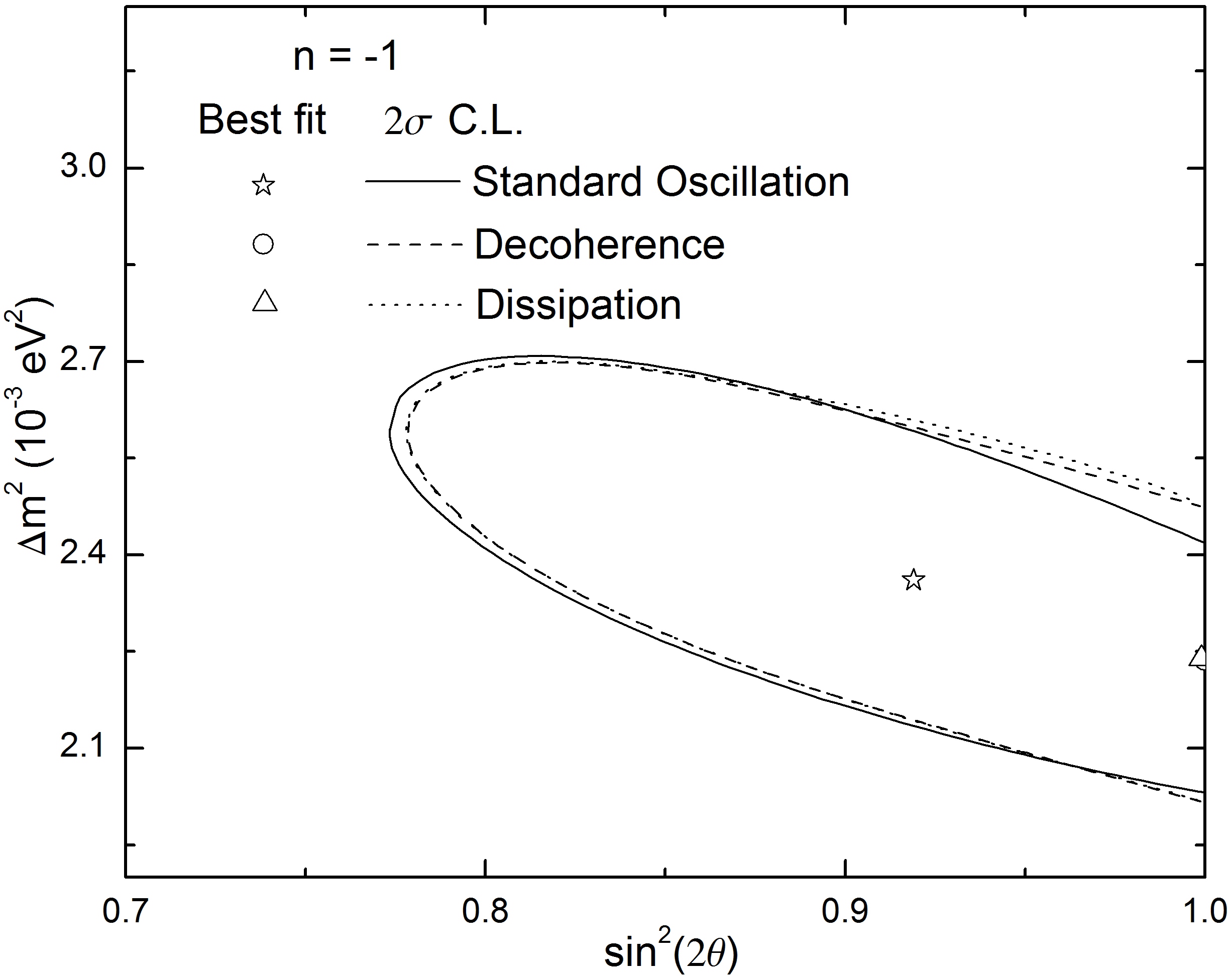}
\includegraphics[width= 5.3 cm, height=5.62cm]{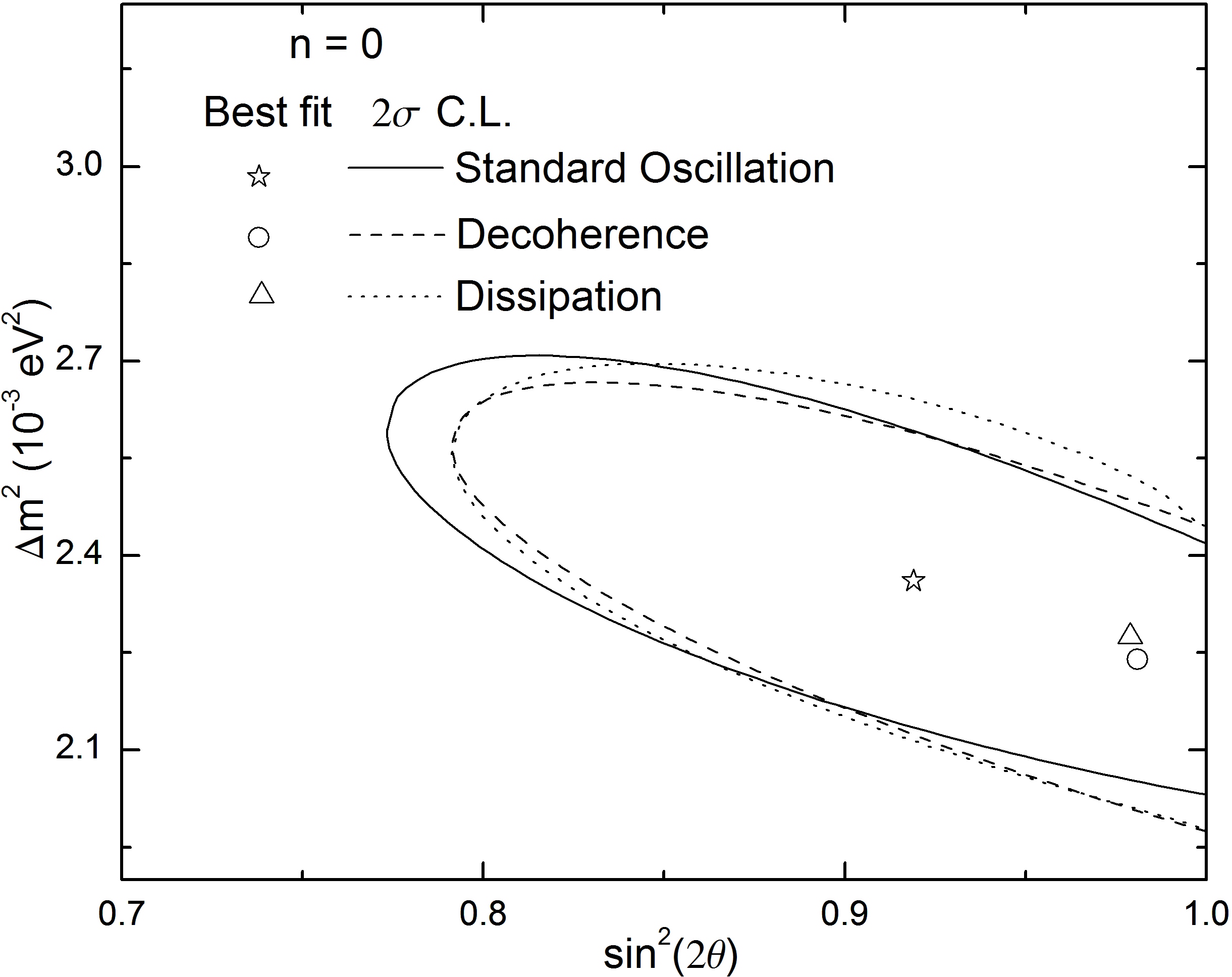}
\includegraphics[width= 5.3 cm, height=5.62cm]{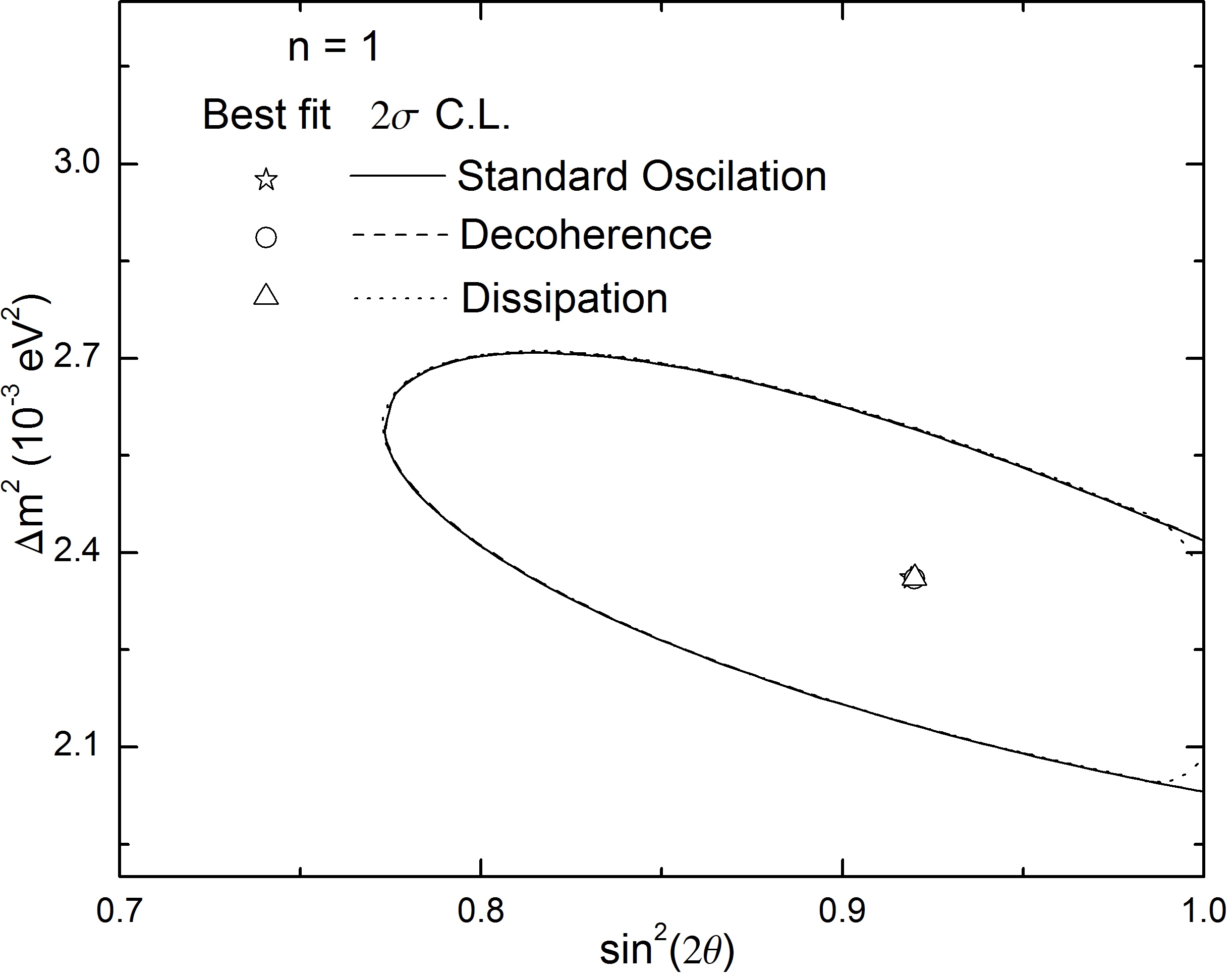}
\includegraphics[width= 5.3 cm, height=5.62cm]{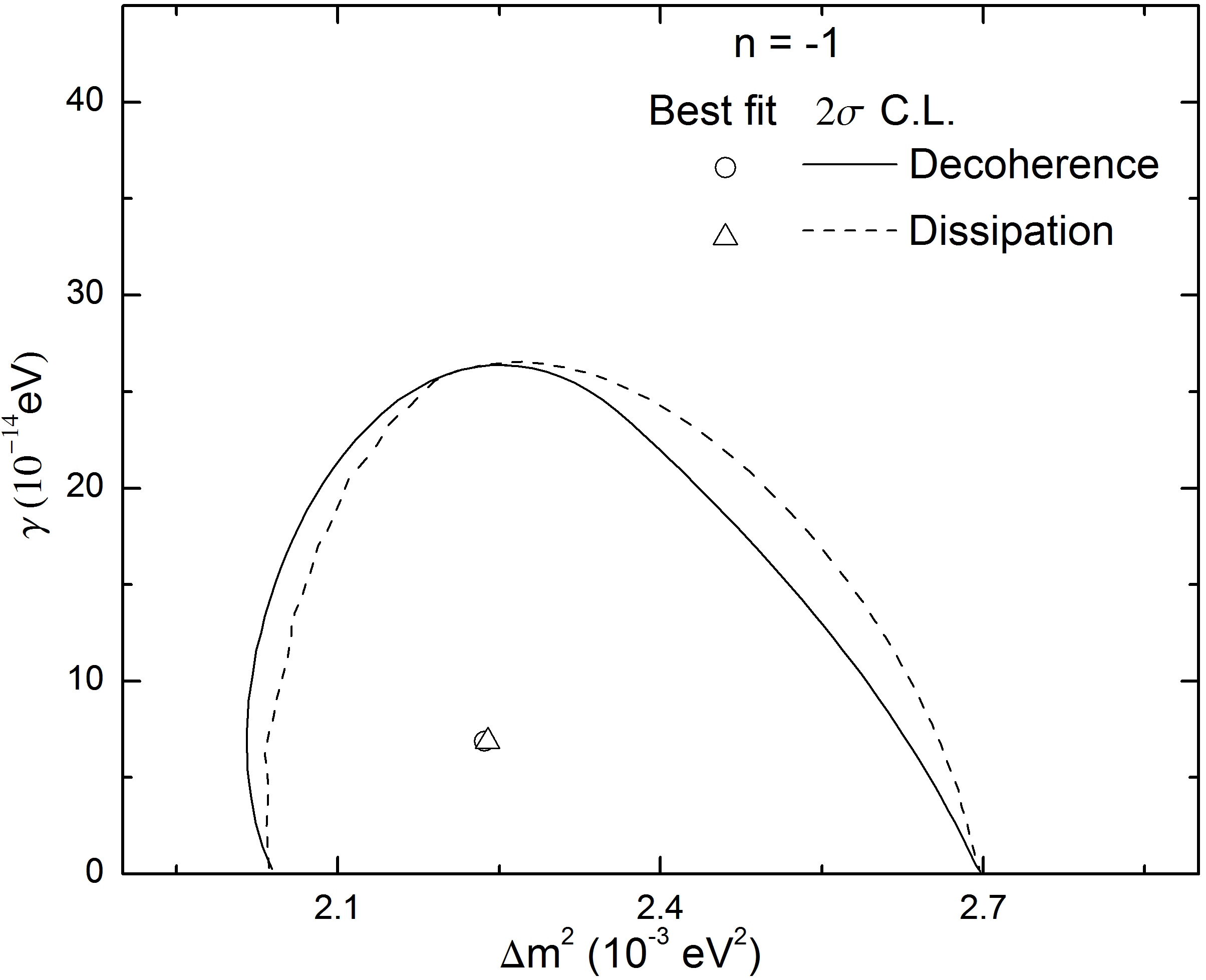}
\includegraphics[width= 5.2 cm, height=5.62cm]{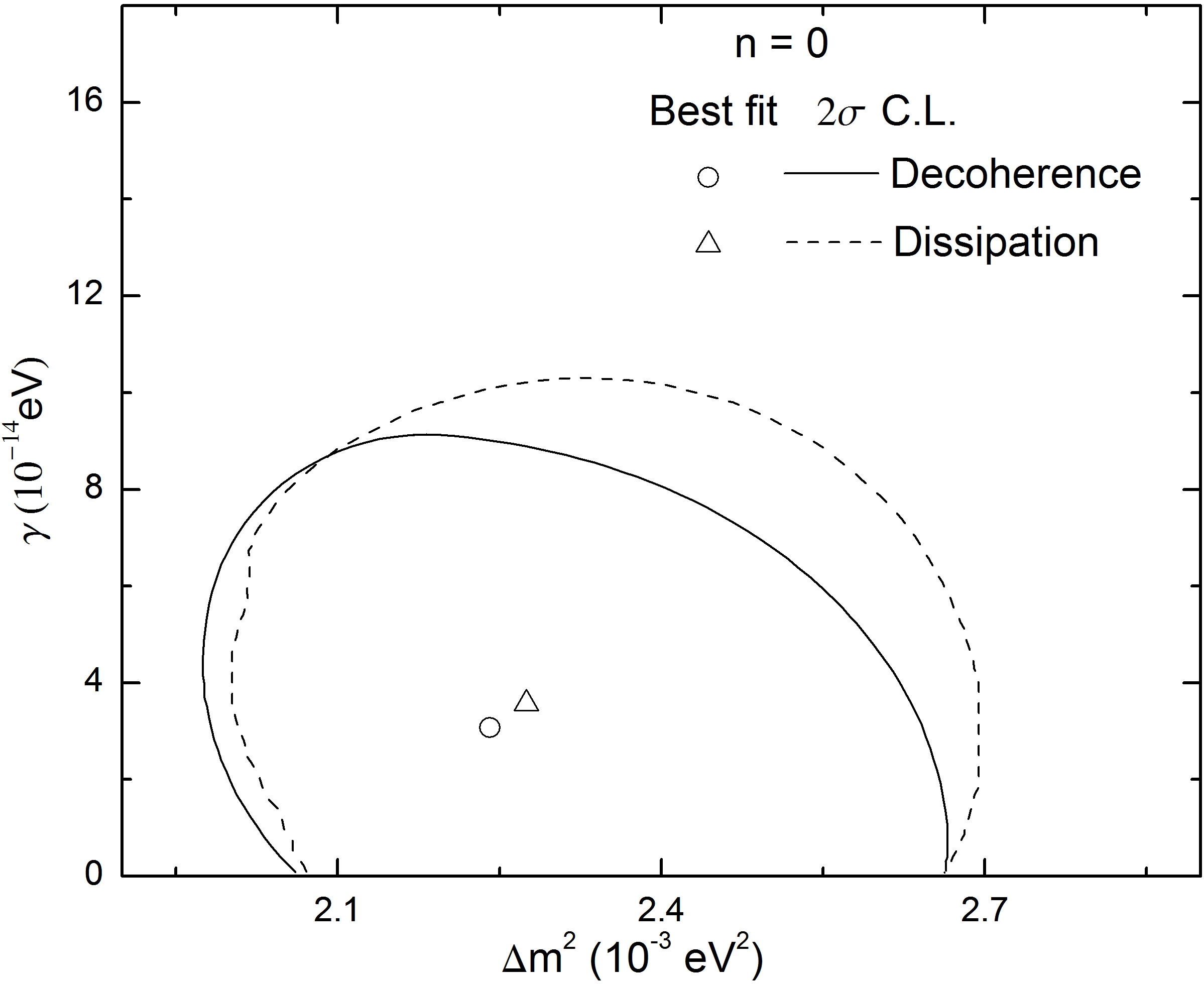}
\includegraphics[width= 5.3 cm, height=5.62cm]{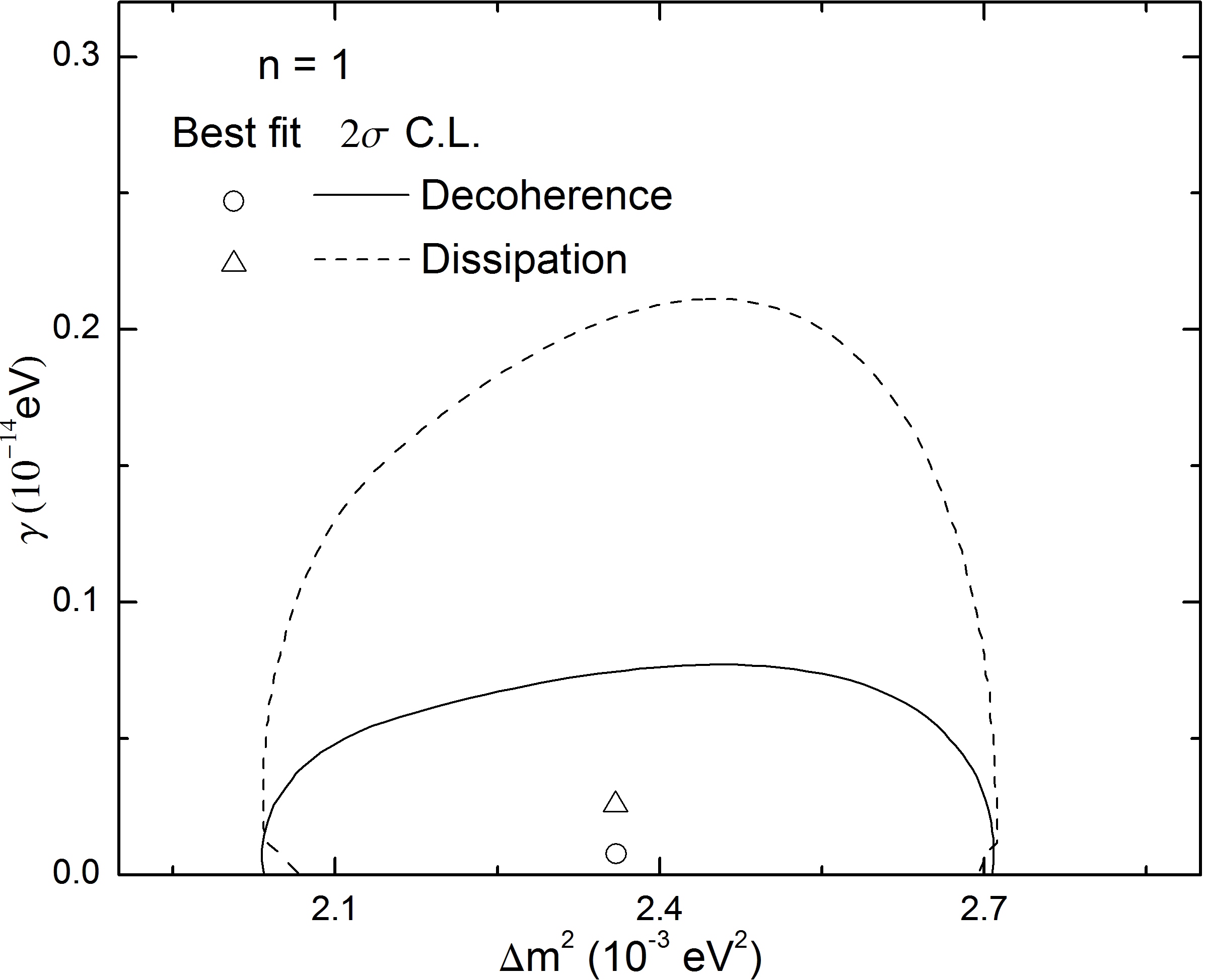}
\includegraphics[width= 5.4 cm, height=5.62cm]{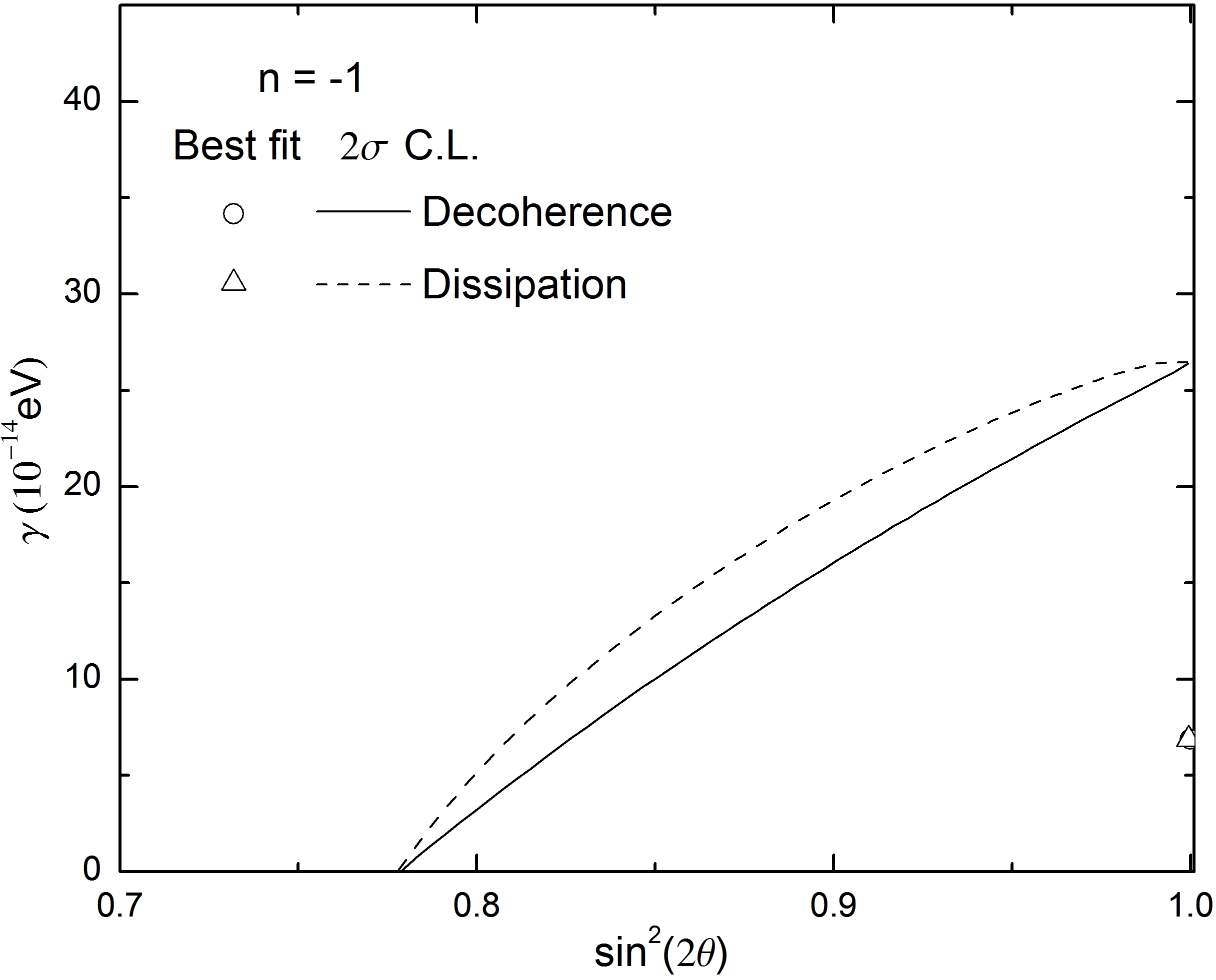}
\includegraphics[width= 5.3 cm, height=5.62cm]{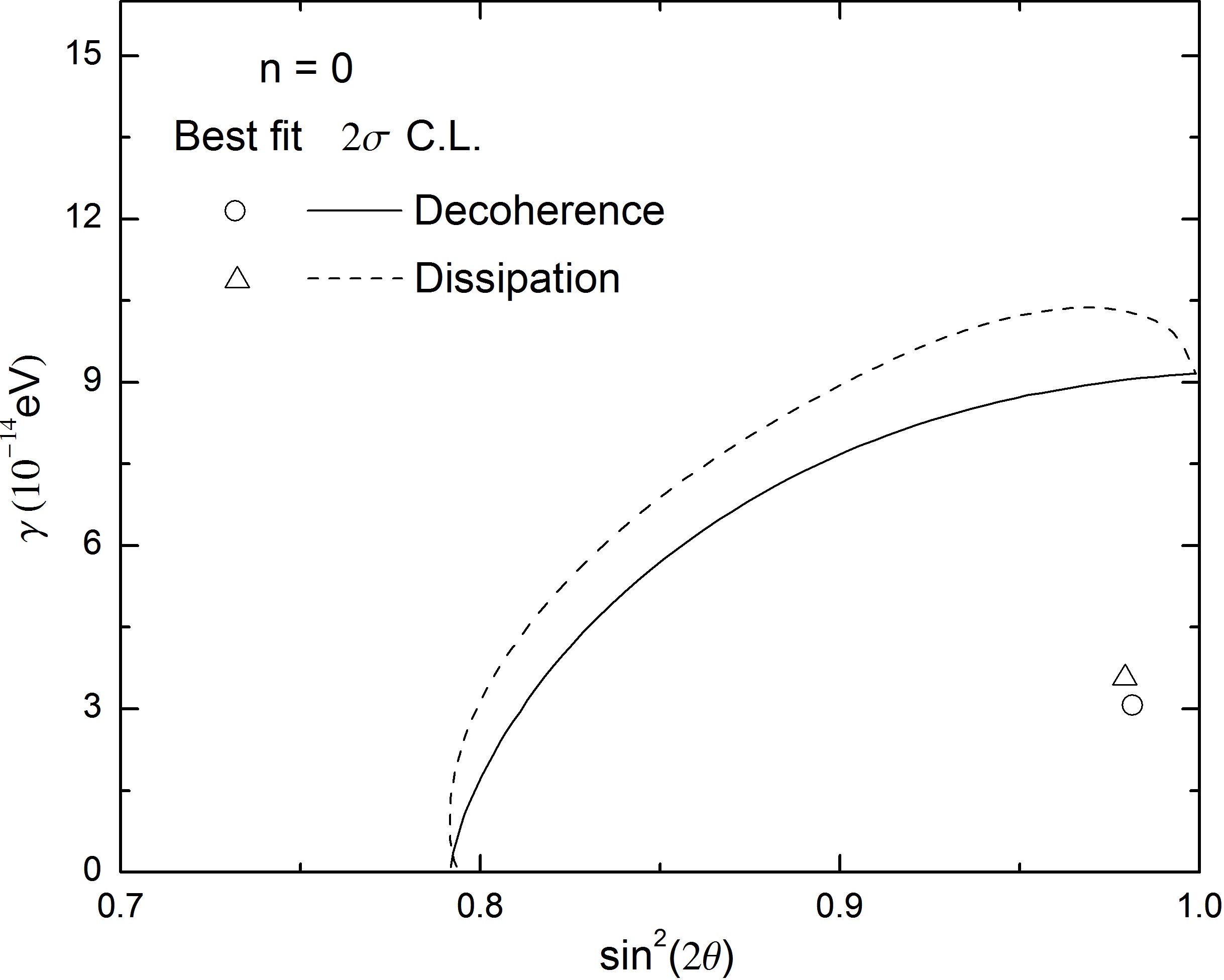}
\includegraphics[width= 5.3 cm, height=5.62cm]{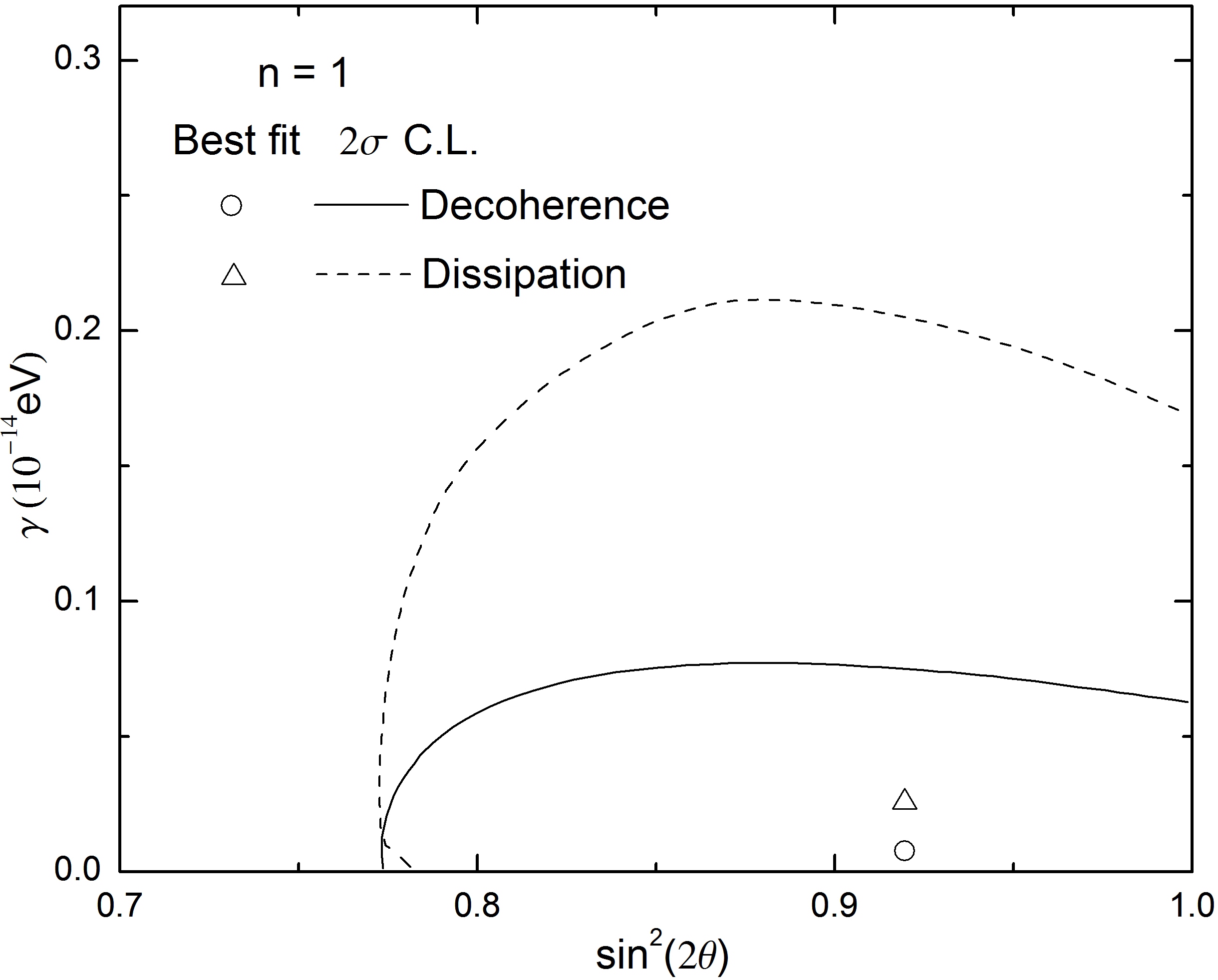}
\caption{Contours at $95\%$ C. L. and best fits obtained from the three models studied. Top: contours with the regions allowed to standard oscillation parameters. In this case, there are three contours for each dissipative models with $n=0, \pm1$ and the same contour for the standard oscillation model. Middle: Limits on $\gamma$ as a function of $\Delta m^{2}$ for the cases  $n=0, \pm1$. Bottom: Limits on $\gamma$ as a function of $\sin^{2}(2\theta)$ considering too $n=0, \pm1$.}
\label{fig.iii}
\end{figure*}

In order to clear up the differences between the dissipative models, we analyze three configurations, $n=0, \pm 1$ on dissipative parameter in each model. The Fig. \ref{fig.iii} shows the best fit values and contours at $95\%$ C.L. for each pair of parameters. At the top in Fig.\ref{fig.iii} are the contours for standard oscillation parameters. We can see that the regions are different from each other due to dissipative effect intensity that depending on $n$ value. When $n=-1$ the standard oscillation model best fit is different from the \textit{Case 1} and \textit{Case 7} models which have the same best fit. To $n=0$, the best fit of the dissipative models  tends to the standard oscillation model best fit. Finally, when $n=1$ the dissipative effect becomes very weak and the three best fits, standard oscillation model and dissipative models, are equal.

\begin{figure*}[!htb]
\center
\includegraphics[width= 8.1 cm, height=7.5cm]{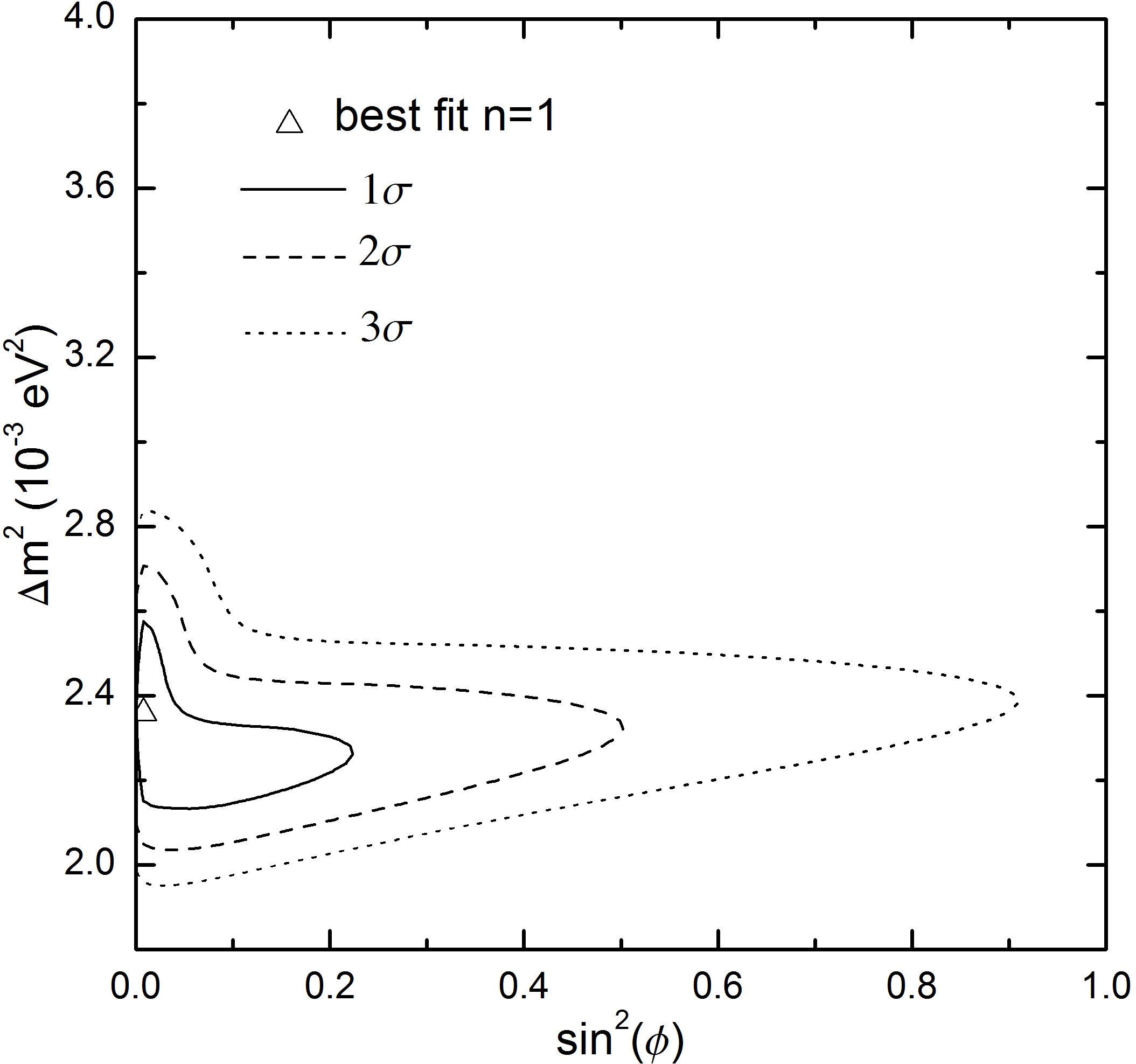}
\includegraphics[width= 8.1 cm, height=7.5cm]{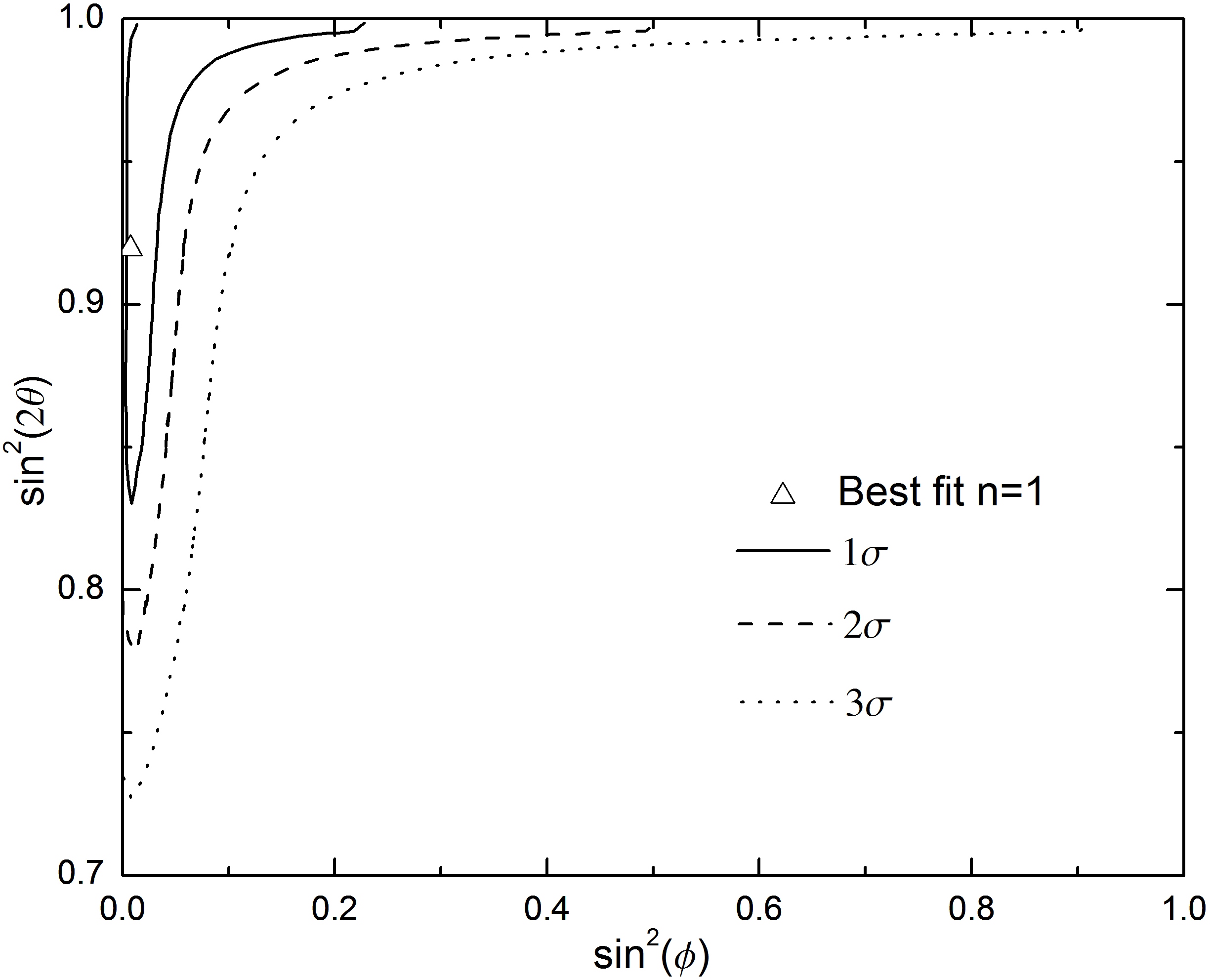}
\caption{The contours obtained when $\gamma$ is fixed in its best fit value. On the left, $\Delta m^{2}$ is shown as function of $\sin^{2}\phi$ and on the right, $\sin^{2}(2\theta)$ as function of $\sin^{2}\phi$. In the two situations there are contours with $1, 2$ and $3 \sigma$ of confidence level.}
\label{fig.iiii}
\end{figure*}

In this approach, the energy dependence on $\gamma$, given by Eq. (\ref{i.iii}), has an important role and each model changes the limits of $\Delta m^{2}$ and $\sin^{2}(2\theta)$ when $n$ varies. This is possible to see in the middle and bottom in Fig. \ref{fig.iii}. Interesting enough, when $n=0$ the models impose on $\sin^{2}(2\theta)$  a stronger limit than when $n=-1, 1$. When $n\geq1$, $\gamma$ must be small, then, we expect that the dissipative effect becomes weak and effectively  less important and in this case the dissipative models tend to the standard oscillation model. It is important to note that when $n=-1, 0$ there are regions outside the standard oscillation region of at $95\%$ C.L.. This can be seen in the top of Fig. \ref{fig.iii}, but when $n=1$, the model \textit{Case 7} has the smaller region than the other models.

\section{The CP Phase}

The \textit{Case 7} model has a new and important difference from the other models. \textit{Case 7} model has a Majorana CP phase in survival probability and in the Table \ref{tab:table2} the value of this phase is non-zero in the global case. Indeed, the best fit to this new parameter in the global case with $n=0, -1$ is maximum, $\sin^{2}(\phi)=1$, and $\sin^{2}(\phi)=0.01$ when $n=1$. Notice that $\sin^{2}(\phi)=0$ when $n=-1, 0$ and $\sin^{2}(\phi)=0.10$ when $n=1$ in individual analysis from neutrinos and antineutrinos. Therefore, the CP phase seems to be an additive parameter that has an important role and its consequences are very interesting.  

The value obtained to CP phase in the global analysis makes the survival probability in Eq. (\ref{i.ii}) to be different when we treat neutrinos or antineutrinos. CP violation can be occur in neutrinos oscillation when we use an open quantum systems approach. Although our analysis did not find sensibility for this parameter, we investigate the CP phase when the $\gamma$ is fixed in its best fit value and thus, we get the behavior the CP phase as function of $\Delta m^{2}$ and $\sin^{2}(2\theta)$. 

The case where $n=1$ was the only one that showed some sensitivity to $\phi$ and in Fig. \ref{fig.iiii} the contours obtained when $\gamma$ is fixed in the best fit value. The limits on $\Delta m^{2}$ $(\sin^{2}(2\theta))$ and $\sin^{2}(\phi)$ in this situation appear on the left (right) in Fig. \ref{fig.iiii}. The $2 \sigma$ region shows that $\sin^{2}(\phi)<0.5$ and the $\Delta m^{2}$ $(\sin^{2}(2\theta))$ limit is inside the same region obtained to standard oscillation model, as it is possible to see at the top right in Fig.\ref{fig.iii}. It is possible to see also that when $\sin^{2}(\phi)\rightarrow 0$ the limit on $\Delta m^{2}$ becomes different from the usual and this explains why the contour obtained with \textit{Case 7} model is smaller than the standard oscillation model contour.

On the other hand, if we take the value of $\gamma$ at $2\sigma$ C.L., $\gamma=0.14\times 10^{-14}$ eV, then $\sin^{2}(\phi)\sim 0$\footnote{The exact value is $\sin^{2}(\phi)= 0.003$ or $\phi=0.06$ rad and in this situation $\chi^{2}/dof=1.08$.} and CP violation in this condiction can be negligible because in the last term of Eq.(\ref{i.ii}) tends strongly to zero.

The analysis performed for \textit{Case 7} model indicates that CP violation can appear even in two-neutrino oscillations. This CP violation has an import consequence once that this approach violates the temporal symmetry \cite{ben, workneut, pet,dav,len, lin}. In fact, the addition of the CP violation in open quantum system approach, that already violates the temporal symmetry, composes an unusual CPT violation, since it occurs even considering neutrinos equivalent to antineutrinos. 

However, it is important to have in mind that the dissipative models contain the usual oscillation parameter and comparing the $\Delta \chi^{2}$ between the dissipative and standard oscillation patterns the biggest difference is $2.4$ and, therefore, these dissipative models are not statistically favored. We have calculated the \textit{p}-value for the standard model in global case and we find $45.87\%$ while the \textit{p}-value to the best dissipative model in global case, when $n=0$, is $47.60\%$. So, we must conclude that the results obtained with all dissipative models do not have statistical preference and, then, we can keep the focus in the limits to dissipative effects, as well as to Majorana CP phase value.


\section{Comments and Conclusion}

We have presented a simple data analysis from MINOS experiment using the open quantum system approach, where the survival probabilities take into account the dissipative effects adding only one parameter in the theory \cite{workneut}. We test our simple approach considering the standard oscillation model in order to verify if the obtained results are suitable to understand the current MINOS result. Our results showed good agreement with MINOS Collaboration results, both for neutrino and antineutrinos \cite{mimi1,mimi4}. 

After this, we performed the analysis using the open quantum system approach where dissipative effects are added to the oscillation phenomena. Two specific models were analyzed, but each dissipative model was analyzed in five different conditions, once that a power-law exponential has been imposed on dissipative parameter. 

The first models, \textit{Case 1}, added only decoherence like dissipative effect in standard oscillation model and the second model, \textit{Case 7}, considers an original condition on dissipative effects. It leads to a most general effect that includes also decoherence and other dissipative effects \cite{workneut}. 


We performed the analysis for neutrinos and antineutrinos and due to consistecy in our results, we imposed equivalence between neutrinos and antineutrinos and perform the global analysis focusing in the cases where $n=0,\pm1$ in power-law of the $\gamma_{0}$ parameter.

 

The results obtained with global hypothesis showed that the oscillation model fits very well the MINOS data. Dissipative effects have low contribution and statistically negligible, although these models present rich phenomenology to be studied. In particular, with the \textit{Case 7} model we obtained a limit to dissipative effects and the Majorana CP phase can have non-zero values in the three possibilities where $n=0, \pm1$. Then, this model, even in two neutrino oscillation, can present CP violation.  Interesting enough, when we treat neutrino and antineutrino separately, the Majorana CP phase is zero in most part of the cases, but to fit the global hypothesis, we find a non-zero CP phase.


In special, we detail the situation where $n=1$ in \textit{Case 7} model and although the dissipative effects are less effective here, the results are interesting. As it can be seen, the \textit{Case 7} model presents effects that can be described by mean of the Majorana CP phase only. When we fixed the $\gamma$ in the best fit the sensitivity in relation to Majorana CP phase becomes significant and as it is shown in Fig.\ref{fig.iiii}. The CP phase is responsible by reduction of the contour region on the top right of Fig.\ref{fig.iii}. However, we point out that in open quantum system approach the temporal symmetry is violated and together with CP violation result we arrive in an unusual CPT violation that is different from the usual CPT symmetry.

In summary, the open quantum system is a rich approach that can include many interesting effects and possibilities of study. Here, we applied this theory in MINOS data analysis and we investigate some intriguing results. The dissipative effects can lead us to new phenomena and consequences. In this work, for example, the Majorana CP phase is kept even in two neutrino oscillation.

\begin{acknowledgments}
The authors thank to the Brazilian Agencies FAPESP (grant 2012/00857-6) and CNPq for several financial supports. 
\end{acknowledgments}






\begin{thebibliography}{29}
\expandafter\ifx\csname natexlab\endcsname\relax\def\natexlab#1{#1}\fi
\expandafter\ifx\csname bibnamefont\endcsname\relax
  \def\bibnamefont#1{#1}\fi
\expandafter\ifx\csname bibfnamefont\endcsname\relax
  \def\bibfnamefont#1{#1}\fi
\expandafter\ifx\csname citenamefont\endcsname\relax
  \def\citenamefont#1{#1}\fi
\expandafter\ifx\csname url\endcsname\relax
  \def\url#1{\texttt{#1}}\fi
\expandafter\ifx\csname urlprefix\endcsname\relax\def\urlprefix{URL }\fi
\providecommand{\bibinfo}[2]{#2}
\providecommand{\eprint}[2][]{\url{#2}}

\bibitem[{\citenamefont{Oliveira and Guzzo}(2010)}]{workneut}
\bibinfo{author}{\bibfnamefont{R.~L.~N.} \bibnamefont{Oliveira}}
  \bibnamefont{and} \bibinfo{author}{\bibfnamefont{M.~M.} \bibnamefont{Guzzo}},
  \bibinfo{journal}{Eur. Phys. Jour. C} \textbf{\bibinfo{volume}{69}},
  \bibinfo{pages}{493} (\bibinfo{year}{2010}).

\bibitem[{\citenamefont{Benatti and Floreanini}(2000)}]{ben}
\bibinfo{author}{\bibfnamefont{F.}~\bibnamefont{Benatti}} \bibnamefont{and}
  \bibinfo{author}{\bibfnamefont{R.}~\bibnamefont{Floreanini}},
  \bibinfo{journal}{JHEP} \textbf{\bibinfo{volume}{02}}, \bibinfo{pages}{32}
  (\bibinfo{year}{2000}).

\bibitem[{\citenamefont{Lindblad}(1976)}]{lin}
\bibinfo{author}{\bibfnamefont{G.}~\bibnamefont{Lindblad}},
  \bibinfo{journal}{Commun. Phys.} \textbf{\bibinfo{volume}{48}},
  \bibinfo{pages}{119} (\bibinfo{year}{1976}).

\bibitem[{\citenamefont{Gorini and Kossakowski}(1976)}]{gor}
\bibinfo{author}{\bibfnamefont{V.}~\bibnamefont{Gorini}} \bibnamefont{and}
  \bibinfo{author}{\bibfnamefont{A.}~\bibnamefont{Kossakowski}},
  \bibinfo{journal}{J. Math. Phys.} \textbf{\bibinfo{volume}{17}},
  \bibinfo{pages}{821} (\bibinfo{year}{1976}).

\bibitem[{\citenamefont{Davies}(1974)}]{dav}
\bibinfo{author}{\bibfnamefont{E.~B.} \bibnamefont{Davies}},
  \bibinfo{journal}{Commun. Phys.} \textbf{\bibinfo{volume}{39}},
  \bibinfo{pages}{91} (\bibinfo{year}{1974}).

\bibitem[{\citenamefont{Alicki and Lendi}(1987)}]{len}
\bibinfo{author}{\bibfnamefont{R.}~\bibnamefont{Alicki}} \bibnamefont{and}
  \bibinfo{author}{\bibfnamefont{K.}~\bibnamefont{Lendi}},
  \emph{\bibinfo{title}{Quantum dynamical semigrups and applications, Lect.
  Notes Phys.}} (\bibinfo{publisher}{Springer-Verlag},
  \bibinfo{address}{Berlim}, \bibinfo{year}{1987}).

\bibitem[{\citenamefont{An et~al.}(2012)}]{day}
\bibinfo{author}{\bibfnamefont{F.~P.} \bibnamefont{An}} \bibnamefont{et~al.}
  (\bibinfo{collaboration}{Daya Bay Collaboration}), \bibinfo{journal}{Phys.
  Rev. Lett.} \textbf{\bibinfo{volume}{108}}, \bibinfo{pages}{171803}
  (\bibinfo{year}{2012}).

\bibitem[{\citenamefont{Abe et~al.}(2012)}]{double}
\bibinfo{author}{\bibfnamefont{Y.}~\bibnamefont{Abe}} \bibnamefont{et~al.}
  (\bibinfo{collaboration}{Double Chooz Collaboration}),
  \bibinfo{journal}{Phys. Rev. Lett.} \textbf{\bibinfo{volume}{108}},
  \bibinfo{pages}{131801} (\bibinfo{year}{2012}).

\bibitem[{\citenamefont{Adamson et~al.}(2011{\natexlab{a}})}]{mimi2}
\bibinfo{author}{\bibfnamefont{P.}~\bibnamefont{Adamson}} \bibnamefont{et~al.}
  (\bibinfo{collaboration}{MINOS Collaboration}), \bibinfo{journal}{Phys. Rev.
  Lett.} \textbf{\bibinfo{volume}{107}}, \bibinfo{pages}{021801}
  (\bibinfo{year}{2011}{\natexlab{a}}).

\bibitem[{\citenamefont{Adamson et~al.}(2012{\natexlab{a}})}]{mimi3}
\bibinfo{author}{\bibfnamefont{P.}~\bibnamefont{Adamson}} \bibnamefont{et~al.}
  (\bibinfo{collaboration}{MINOS Collaboration}), \bibinfo{journal}{Phys.Rev.D}
  \textbf{\bibinfo{volume}{86}}, \bibinfo{pages}{052007}
  (\bibinfo{year}{2012}{\natexlab{a}}).

\bibitem[{\citenamefont{Adamson et~al.}(2006)}]{1min}
\bibinfo{author}{\bibfnamefont{P.}~\bibnamefont{Adamson}} \bibnamefont{et~al.}
  (\bibinfo{collaboration}{MINOS Collaboration}), \bibinfo{journal}{Phys. Rev.
  Lett} \textbf{\bibinfo{volume}{97}}, \bibinfo{pages}{191801}
  (\bibinfo{year}{2006}).

\bibitem[{\citenamefont{Adamson et~al.}(2011{\natexlab{b}})}]{mimi1}
\bibinfo{author}{\bibfnamefont{P.}~\bibnamefont{Adamson}} \bibnamefont{et~al.}
  (\bibinfo{collaboration}{MINOS Collaboration}), \bibinfo{journal}{Phys. Rev.
  Lett.} \textbf{\bibinfo{volume}{106}}, \bibinfo{pages}{181801}
  (\bibinfo{year}{2011}{\natexlab{b}}).

\bibitem[{\citenamefont{Morgan et~al.}(2006)}]{dea}
\bibinfo{author}{\bibfnamefont{D.}~\bibnamefont{Morgan}} \bibnamefont{et~al.},
  \bibinfo{journal}{Astrop. Phys.} \textbf{\bibinfo{volume}{25}},
  \bibinfo{pages}{311} (\bibinfo{year}{2006}).

\bibitem[{\citenamefont{Lisi et~al.}(2000)\citenamefont{Lisi, Marrone, and
  Montanino}}]{lis}
\bibinfo{author}{\bibfnamefont{E.}~\bibnamefont{Lisi}},
  \bibinfo{author}{\bibfnamefont{A.}~\bibnamefont{Marrone}}, \bibnamefont{and}
  \bibinfo{author}{\bibfnamefont{D.}~\bibnamefont{Montanino}},
  \bibinfo{journal}{Phys. Rev. Lett.} \textbf{\bibinfo{volume}{85}},
  \bibinfo{pages}{1166} (\bibinfo{year}{2000}).

\bibitem[{\citenamefont{Fogli et~al.}(2007)\citenamefont{Fogli, Lisi, Marrone,
  Montanino, and A.Palazzo}}]{fo}
\bibinfo{author}{\bibfnamefont{G.~L.} \bibnamefont{Fogli}},
  \bibinfo{author}{\bibfnamefont{E.}~\bibnamefont{Lisi}},
  \bibinfo{author}{\bibfnamefont{A.}~\bibnamefont{Marrone}},
  \bibinfo{author}{\bibfnamefont{D.}~\bibnamefont{Montanino}},
  \bibnamefont{and} \bibinfo{author}{\bibnamefont{A.Palazzo}},
  \bibinfo{journal}{Phys. Rev. D} \textbf{\bibinfo{volume}{76}},
  \bibinfo{pages}{033006} (\bibinfo{year}{2007}).

\bibitem[{\citenamefont{Fogli et~al.}(2003)\citenamefont{Fogli, Lisi, Marrone,
  and Montanino}}]{liss}
\bibinfo{author}{\bibfnamefont{G.~L.} \bibnamefont{Fogli}},
  \bibinfo{author}{\bibfnamefont{E.}~\bibnamefont{Lisi}},
  \bibinfo{author}{\bibfnamefont{A.}~\bibnamefont{Marrone}}, \bibnamefont{and}
  \bibinfo{author}{\bibfnamefont{D.}~\bibnamefont{Montanino}},
  \bibinfo{journal}{Phys. Rev. D} \textbf{\bibinfo{volume}{67}},
  \bibinfo{pages}{093006} (\bibinfo{year}{2003}).

\bibitem[{\citenamefont{Hooper et~al.}(1995)}]{dan}
\bibinfo{author}{\bibfnamefont{D.}~\bibnamefont{Hooper}} \bibnamefont{et~al.},
  \bibinfo{journal}{Phys. Lett. B} \textbf{\bibinfo{volume}{75}},
  \bibinfo{pages}{2650} (\bibinfo{year}{1995}).

\bibitem[{\citenamefont{Farzan et~al.}(2008)\citenamefont{Farzan, Schwetz, and
  Smirnov}}]{yu}
\bibinfo{author}{\bibfnamefont{Y.}~\bibnamefont{Farzan}},
  \bibinfo{author}{\bibfnamefont{T.}~\bibnamefont{Schwetz}}, \bibnamefont{and}
  \bibinfo{author}{\bibfnamefont{A.~Y.} \bibnamefont{Smirnov}},
  \bibinfo{journal}{JHEP} \textbf{\bibinfo{volume}{67}}, \bibinfo{pages}{0807}
  (\bibinfo{year}{2008}).

\bibitem[{\citenamefont{Ohlsson}(2001)}]{mmy}
\bibinfo{author}{\bibfnamefont{T.}~\bibnamefont{Ohlsson}},
  \bibinfo{journal}{Phys. Lett. B} \textbf{\bibinfo{volume}{502}},
  \bibinfo{pages}{159} (\bibinfo{year}{2001}).

\bibitem[{\citenamefont{Gago et~al.}(2001)}]{fun1}
\bibinfo{author}{\bibfnamefont{A.~M.} \bibnamefont{Gago}} \bibnamefont{et~al.},
  \bibinfo{journal}{Phys. Rev. D} \textbf{\bibinfo{volume}{63}},
  \bibinfo{pages}{073001} (\bibinfo{year}{2001}).

\bibitem[{\citenamefont{Barenboim and Mavromato}(2005)}]{gab}
\bibinfo{author}{\bibfnamefont{G.}~\bibnamefont{Barenboim}} \bibnamefont{and}
  \bibinfo{author}{\bibfnamefont{N.~E.} \bibnamefont{Mavromato}},
  \bibinfo{journal}{JHEP} \textbf{\bibinfo{volume}{01}}, \bibinfo{pages}{31}
  (\bibinfo{year}{2005}).

\bibitem[{\citenamefont{Liu et~al.}(2011)}]{chun}
\bibinfo{author}{\bibfnamefont{C.}~\bibnamefont{Liu}} \bibnamefont{et~al.},
  \bibinfo{journal}{Phys.Lett.B} \textbf{\bibinfo{volume}{702}},
  \bibinfo{pages}{154} (\bibinfo{year}{2011}).

\bibitem[{\citenamefont{Ellis et~al.}(1984)}]{ell}
\bibinfo{author}{\bibfnamefont{J.}~\bibnamefont{Ellis}} \bibnamefont{et~al.},
  \bibinfo{journal}{Nucl. Phys. B} \textbf{\bibinfo{volume}{241}},
  \bibinfo{pages}{381} (\bibinfo{year}{1984}).

\bibitem[{\citenamefont{Bardeen et~al.}(1973)\citenamefont{Bardeen, Carter, and
  Hawking}}]{haw}
\bibinfo{author}{\bibfnamefont{J.~H.} \bibnamefont{Bardeen}},
  \bibinfo{author}{\bibfnamefont{B.}~\bibnamefont{Carter}}, \bibnamefont{and}
  \bibinfo{author}{\bibfnamefont{S.}~\bibnamefont{Hawking}},
  \bibinfo{journal}{Commun. Math. Phys.} \textbf{\bibinfo{volume}{31}},
  \bibinfo{pages}{161} (\bibinfo{year}{1973}).

\bibitem[{\citenamefont{Breuer and Petruccione}(2002)}]{pet}
\bibinfo{author}{\bibfnamefont{H.~P.} \bibnamefont{Breuer}} \bibnamefont{and}
  \bibinfo{author}{\bibfnamefont{F.}~\bibnamefont{Petruccione}},
  \emph{\bibinfo{title}{The Theory of Open Quantum Systems, Lect. Notes Phys.}}
  (\bibinfo{publisher}{Oxford University Press}, \bibinfo{address}{Oxford},
  \bibinfo{year}{2002}).

\bibitem[{\citenamefont{Weiss}(1993)}]{uri}
\bibinfo{author}{\bibfnamefont{U.}~\bibnamefont{Weiss}},
  \emph{\bibinfo{title}{Quantum Dissipative Systems}}, vol.
  \bibinfo{volume}{XIII} (\bibinfo{publisher}{World ScientificHermann},
  \bibinfo{address}{Singapore}, \bibinfo{year}{1993}).

\bibitem[{\citenamefont{Joos et~al.}(2003)}]{joo}
\bibinfo{author}{\bibfnamefont{E.}~\bibnamefont{Joos}} \bibnamefont{et~al.},
  \emph{\bibinfo{title}{it Decoherence and the Appearance of Classical World in
  Qunatum Theoriy}} (\bibinfo{publisher}{2ed. Springer}, \bibinfo{address}{New
  York}, \bibinfo{year}{2003}).

\bibitem[{\citenamefont{Giunti and Kim}(2007)}]{kim}
\bibinfo{author}{\bibfnamefont{C.}~\bibnamefont{Giunti}} \bibnamefont{and}
  \bibinfo{author}{\bibfnamefont{C.~W.} \bibnamefont{Kim}},
  \emph{\bibinfo{title}{Fundamentals of neutrino physics and astrophysics}}
  (\bibinfo{publisher}{Oxford University Press}, \bibinfo{address}{New York},
  \bibinfo{year}{2007}).

\bibitem[{\citenamefont{Adamson et~al.}(2012{\natexlab{b}})}]{mimi4}
\bibinfo{author}{\bibfnamefont{P.}~\bibnamefont{Adamson}} \bibnamefont{et~al.}
  (\bibinfo{collaboration}{MINOS Collaboration}), \bibinfo{journal}{Phys. Rev.
  Lett.} \textbf{\bibinfo{volume}{108}}, \bibinfo{pages}{191801}
  (\bibinfo{year}{2012}{\natexlab{b}}).

\end{thebibliography}

\end{document}